\documentclass[fleqn,usenatbib]{mnras}

\usepackage{newtxtext,newtxmath}

\usepackage[T1]{fontenc}
\usepackage{ae,aecompl}
\usepackage[acronym]{glossaries}

\usepackage{graphicx}	
\usepackage{amsmath}	
\usepackage{amssymb}	
\usepackage{epsfig}

\title[21cm  Global Signal Extraction]{ Extracting the 21cm  Global Signal using Artificial Neural Networks}

\author[M. Choudhury et al]{
Madhurima Choudhury,$^{1}$\thanks{E-mail: madhurimachoudhury811@gmail.com}
Abhirup Datta,$^{1}$\thanks{E-mail: abhirup.datta@iiti.ac.in} and 
Arnab Chakraborty,$^{1}$ 
\\
$^{1}$Discipline of Astronomy, Astrophysics and Space Engineering, Indian Institute of Technology Indore, 453552, India.\\
}

\date{Accepted XXX. Received YYY; in original form ZZZ}

\pubyear{2017}

\newacronym{ann}{ANN}{Artificial Neural Networks}

\begin{document}
\label{firstpage}
\pagerange{\pageref{firstpage}--\pageref{lastpage}}
\maketitle

\begin{abstract}
 The study of the cosmic Dark Ages, Cosmic Dawn, and Epoch of Reionization (EoR) using the all-sky averaged redshifted HI 21cm signal, are some of the key science goals of most of the ongoing or upcoming experiments, for example, EDGES, SARAS, and the SKA. This signal can be detected by averaging over the entire sky,  using a single radio telescope, in the form of a Global signal as a function of only redshifted HI 21cm frequencies. One of the major challenges faced while detecting this signal is the dominating, bright foreground. The success of such detection lies in the accuracy of the foreground removal. The presence of instrumental gain fluctuations, chromatic primary beam, radio frequency interference (RFI) and the Earth's ionosphere corrupts any observation of radio signals from the Earth. Here, we propose the use of Artificial Neural Networks (ANN) to extract the faint redshifted 21cm Global signal buried in a sea of bright Galactic foregrounds and contaminated by different instrumental models. The most striking advantage of using ANN is the fact that, when the corrupted
signal is fed into a trained network, we can simultaneously extract the signal as well as foreground parameters very accurately. Our results show that ANN can detect the Global signal with $\gtrsim 92 \%$ accuracy even in cases of mock observations where the instrument has some residual time-varying gain across the spectrum. 
\end{abstract}




\section{Introduction}
The redshifted 21cm  line of neutral hydrogen presents a unique probe of the evolution of the neutral intergalactic medium (IGM), from the Cosmic Dark Ages through Cosmic Dawn and Cosmic Reionization \citep{Furlanetto_2006, Morales_2010,Pritchard_2012}. The interplay between the CMB temperature, the kinetic temperature and the spin temperature, along with radiative transfer, lead to very interesting physics of the 21cm  signal evolving over a redshift range. The 21cm  line does not saturate, like the $Ly\alpha$, as at these high redshifts, \citep{Fan_2002,Fan_2006,Mortlock_2011}, the IGM remains somewhat translucent at large neutral fractions \citep{Barkana_2005A} due to the Gunn-Peterson effect. The HI 21cm  observations can be used to study evolution of cosmic structure from the linear regime at high redshift (i.e., density-only evolution), and through the non-linear regime associated with luminous source formation. The 21cm  line as a probe, promises to be the richest of all cosmological datasets: HI measurements are sensitive to structures ranging from very large scales down to the source scale set by the cosmological Jeans mass\citep{Barkana_2005B}. \\
The 21cm  cosmological signal is hidden in a bright sea of foregrounds, which are $\sim10^4$ times higher in magnitude. The foreground needs to be characterised and removed, in order to be able to detect this faint signal. In addition to the strong foregrounds, we have ionospheric distortion, RFI and also the frequency response of instrument which cause the sky as seen by the antenna to vary with time and other factors. All these factors, make extracting of the 21cm Global signal, extremely challenging. The most traditional procedure for extracting the faint cosmological signal is by assuming that the spectrally smooth foreground is well characterized and can be removed from the total signal.
What is left as the residual would contain the signatures from the early phases of the formation of the Universe.
There are theoretical models which explain the evolution of the Global signature with redshift, which essentially depends on the interplay of several crucial parameters. The detection of the 21cm  Global Signal would in many ways complement the detection of the 21cm  power spectrum, and provide meaningful insight about the physical parameters for the evolution of the IGM.  Several upcoming experiments plan to observe the sky-averaged signal: examples are, the Shaped Antenna measurement of the background RAdio Spectrum, SARAS, \citep{Patra_2013,Singh_2017}, the Large-Aperture Experiment to Detect the Dark Ages, LEDA \citep{Greenhill_2012,Price_2018}, SCI-HI \citep{Voytek_2014}, the Broadband Instrument for Global Hydrogen Reionisation Signal, BIGHORNS \citep{Sokolowski_2015}. As there is a considerable amount of contamination of the signal by man-made sources (RFI) and ionosphere for all terrestrial observations, there have been some proposals to go to the far
side of the moon, Dark Ages Radio Explorer (DARE) \citep{Burns_2012,Burns_2017}.
There has been a recent report of detection of a flattened and deep absorption trough at 78MHz, of the sky-averaged radio spectrum by EDGES, Experiment to Detect the Global Epoch of Reionization Signature \citep{Bowman_2018}. The reported absorption amplitude is 0.5K, which is exceedingly high, as compared to the expected amplitude of 150-200 mK. If confirmed, this will give us new insight to the existing physics of the evolution of our Universe.  Following this detection, \citet{Fialkov_2018} have explored a wider range of various possible 21cm signals, varying the properties of the dark matter particles, in addition to varying the astrophysical parameters. In \cite{EwallWice_2018}, models which produce an excess radio background, explaining the large amplitude of the absorption feature in the EDGES detection, have been explored. 
The 21cm Global signal can be parametrized and the parameters can be associated with the physical processes as the Universe evolved. The already existing techniques of parameter estimation, (e.g. \citet{Harker_2012}) relies, to a large extent, on the choice of the statistical prior distribution and does not consider realistic scenarios. However, in the later works of \citet{Harker_2015,Harker_2016, Mirocha_2015}, they have used broad uninformative priors, and were able to recover the signal from synthetic datasets, assuming a perfect instrument and simple foregrounds.  In the work by \citet{Tauscher_2017, Tauscher_2018} another technique has been introduced which uses Singular Value Decomposition (SVD) and produce systematics basis functions specifically suited to different observations of the Global signal. \citet{Bernardi_2016} have used a completely Bayesian formalism, HIBayes, in which the signal is modelled as a Gaussian and the foreground is parametrized as a 7th order polynomial. All these, motivate us to search for another technique for parameter extraction from this contaminated spectrum. Implementation of machine learning algorithms, have shown some encouraging results recently, as demonstrated in \citet{Shimabukuro_2017}, \citet{Schmit_2017} for 21cm  power spectrum analyses. In \citet{Hewitt_2016} they have used Fisher formalism to derive predictions of the constraints that 21cm power spectrum measurements would place on the characteristics of the X-ray sources that heated the IGM at high redshifts. \\
In this paper, we have explored Artificial Neural Networks (ANNs), as an alternate technique for parameter extraction for the 21cm Global signal.
We produce the cosmological signal using $\tanh$ parameterization and the Accelerated Reionization Era Simulations(ARES) \citep{Mirocha_2012}. We then choose a suitable foreground model to represent the bright, dominant foregrounds and create a training dataset for the ANN. This plays a role similar to that of the priors, with the additional advantage of taking into consideration realistic datasets. This technique, might enable us to look for more complicated signals and search for a higher dimensional parameter space where parameter estimation becomes computationally costly. To the best of our knowledge, ANN is computationally cheaper and faster than MCMC for parameter extraction for large parameter spaces. 

We begin with a brief review of the physical processes involved in reionization in \textsection \ref{21cmSignal}. Previous reviews that discuss the HI 21cm signal from cosmic reionization in detail, include \citet{Barkana_2001,Loeb_2001,Furlanetto_2006} and in the context of the Square Kilometer Array (SKA), \citet{Carilli_2004,Koopmans_2015}, \citet{Furlanetto_2004b}. In  \textsection\ref{21cmGS simulation}, we discuss how the 21cm Global signal can be simulated. We also give a short overview of the parameterization of the 21cm Global signal and the previous techniques of signal extraction. In \textsection\ref{foregrounds} and \textsection\ref{instrument}, we describe the foregrounds concerned and the instrument dealt with in such experiments, respectively.
We then explain in detail, the concept of \acrlong{ann}, abbreviated as \acrshort{ann}, in \textsection\ref{ANN-intro} followed by a detailed discussion on the application of \acrshort{ann} in extracting the faint cosmological signal from very bright foregrounds in \textsection\ref{ANN-21cmGS} and \textsection\ref{simulation of21cmGS}. In  \textsection\ref{results}, we present the results obtained by our method. Lastly, we will discuss the advantages of using ANNs as an alternative to the traditional fitting techniques in detail in \textsection\ref{Discussions}. 
\begin{figure}
\includegraphics[width=\columnwidth]{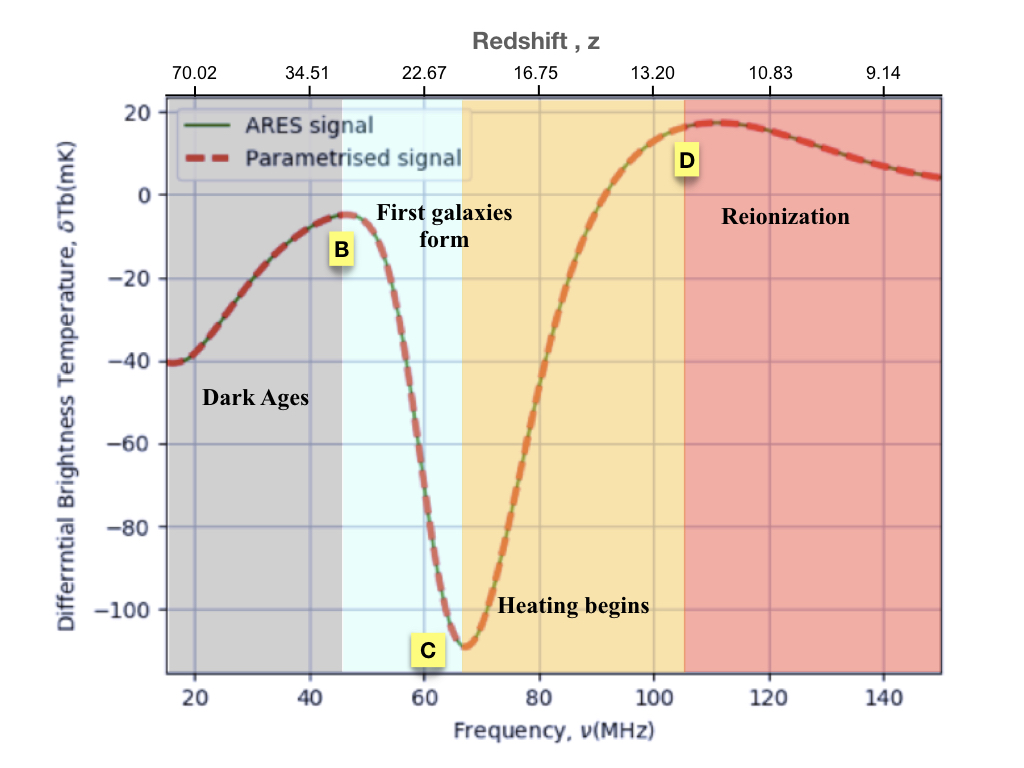}
\caption{The signal constructed using the \textit{tanh} model and the ARES signal.The points B,C,D mark the turning points of the turning point model. Note that the signal using the turning point model is not plotted here.}
\label{fig:global}
\end{figure}
\section{The Cosmological 21cm Signal}\label{21cmSignal}
The hyperfine transition line of atomic hydrogen (in the ground state) arises due to the interaction between the electron and proton spins. The quantity that we measure is the brightness temperature or more accurately called the differential brightness temperature, $\delta T_{b}$, measured relative to the CMB \citep{Pritchard_2015},
\begin{equation}
\delta T_{b} \equiv T_{b} - T_{\gamma}
\end{equation}
\begin{equation}
\begin{split}
\delta T_{b}(\nu) & =\frac{T_{s}-T_{\gamma}}{1+z}(1-\exp^{-\tau_{\nu_{0}}})\\
& \approx 27x_{HI}(1+\delta_{b})\left(\frac{\Omega_{b}h^{2}}{0.023}\right)\left(\frac{0.15}{\Omega_{m,0}h^{2}} \frac{1+z}{10}\right)^{1/2}\\
& ~~\left(1-\frac{T_{\gamma}(z)}{T_{s}}\right)\Big[\frac{\partial_{r} v_{r}}{(1+z)H(z)}\Big]^{-1}
\end{split}
\label{eq:global}
\end{equation}
where, $x_{HI}$ denotes the neutral fraction of hydrogen, $\delta_{b}$ is the fractional over-density of baryons, $\Omega_{b}$ and $\Omega_{M}$ are the baryon and total matter density respectively, in units of the critical density, $H(z)$ is the Hubble parameter and $T_{\gamma}(z)$ is the CMB temperature at redshift z, $T_{s}$ is the spin temperature of neutral hydrogen, and $\partial_{r} v_{r}$ is the velocity gradient along the line of sight. The 21cm Global signal is the sky averaged signal, whose characteristic shape contains information about global cosmic events. The differential brightness temperature can tell us about the ionizing radiation, which destroys neutral hydrogen, the X-rays, which can heat the gas and raise $T_{k}$, and $Ly-\alpha$, which
causes Wouthuysen-Field coupling \citep{Wouthuysen_1952,Field_1959}. Fig.~\ref{fig:global} shows the typical evolution of the signal with frequency. In our calculations, we neglect the peculiar velocity term and the density fluctuation term
in the Global signal (Eqn.~\ref{eq:global}), as it averages out to a linear order and adds to a very small correction. So, the shape of the Global signal broadly depends on the density, neutral fraction and the spin temperature. 
\begin{equation}
   { \delta T_{b}\approx 27(1-x_{i})\left(\frac{\Omega_{b,0}h^{2}}{0.023}\right) \left(\frac{0.15}{\Omega_{m,0}h^{2}} \frac{1+z}{10}\right)^{1/2} \left(1-\frac{T_{\gamma}}{T_{s}}\right)} 
\label{eq:dT_b}
\end{equation}
This equation is primarily used to construct the Global signal and would be the working equation for our work throughout the paper.
The important epochs in the evolution of the signal are labelled in Fig.~\ref{fig:global}. 

The most interesting quantity in the expression for the differential brightness temperature is the spin temperature, $T_{s}$, which  primarily determines the intensity of the 21cm radiation [See Eqn.~\ref{eq:dT_b}]. There are three competing processes that determine $T_{s}$. They are: (1) absorption of CMB photons (as well as stimulated emission); (2) collisions with other hydrogen atoms, free electrons, and protons; and (3) scattering of Lyman alpha photons through excitation and de-excitation.  The spin temperature is defined as \citep{Field_1959, Pritchard_2012}:
\begin{equation}
    T_{s}^{-1}=\frac{T_{\gamma}^{-1}+x_{k}T_{k}^{-1}+x_{\alpha}T_{\alpha}^{-1}}{1+x_{k}+x_{\alpha}},
\label{eq:SpinTemp}
\end{equation}
where, $T_{\gamma}$ is the CMB temperature, $T_{k}$ is the kinetic gas temperature, $T_{\alpha}$ is the temperature related to the existence of ambient Lyman-alpha $(Ly\alpha)$ photons and $x_{c}, x_{\alpha}$ are respectively the collisional coupling and the $Ly\alpha$ coupling terms. In \cite{Barkana_2005A}, the evolution of the Global 21cm signal is discussed in great detail. We briefly explain the evolution of the Global 21cm signal in the following paragraph. \\
At redshifts $z > 200$, free electrons couple $T_{\gamma}$ and $T_{k}$ through Thomson scattering and gas collisions, while the density is high enough to keep $T_{k}$ and $T_{s}$ in equilibrium. Here, $T_{s}= T_{\gamma}$ and there is no 21cm signal. This period is called the \textit{Dark Ages}.
At $z\approx 30-200$, the ionization fraction and density is too low to couple $T_{k}$ to $T_{\gamma}$. Thus, the gas cools adiabatically, with the temperature falling as $(1+z)^2$, which is faster than the rate at which the CMB temperature falls, i.e, as $(1+z)$. Thus the gas becomes colder than the CMB but the mean density is still high enough to provide coupling between $T_{s}$ and $T_{k}$ through collisions. Around this redshift range, the 21cm signal might be seen in absorption against the CMB. At $z\approx 20-30$, collisions can no longer couple $T_{k}$ to $T_{s}$, and thus $T_{s}$ begins to approach CMB. We might expect the first luminous structures near the end of this redshift range. This period is called the \textit{Cosmic Dawn} : the birth of the first luminous sources in the Universe. The $Ly\alpha$ photons from these objects would induce local coupling of $T_{k}$ and $T_{s}$, which might lead to some absorption regions of the 21cm signal. These photons, as well as X-rays from the first luminous sources, could lead to warming of the IGM above the CMB temperature, well before reionization. Hence, one might expect some patches of regions with no signal, absorption and maybe emission,in the 21cm line. 
At $z\approx6-20$ more physical processes come into play. The scattering of the $Ly\alpha$ photons, X-rays from the first galaxies and black holes and weak shocks associated with structure formation warms up the IGM, so that $T_{k}$ is larger than $T_{\gamma}$ \citep{Furlanetto_2004a}. These objects are reionizing the Universe, from the evolution of large scale structure, to a bubble dominated era of HII regions. In this regime, the expected 21cm signal is rich. After reionization, i.e., $z\approx6$, IGM is completely ionized and the 21cm signal is gone again.\\
There are several experiments either proposed or already operational which are designed to detect this Global 21cm signal either in some specific redshift ranges or over the entire range of redshifts ranging from the cosmic \textit{Dark Ages} through \textit{Cosmic Dawn} to \textit{Epoch of Reionization}. In this paper, we consider a large range of redshifts covering most of these three important epochs. 

\section{Simulating the Global Signal}\label{21cmGS simulation}
Based on several theoretical models the cosmological Global 21cm signal changes its shape and amplitude as a function of redshifted frequencies. \citet{Fialkov_2014,Fialkov_2015,Cohen_2017} have used a semi-numerical approach to model possible 21cm Global signals in the redshift range, $z\sim6-40$, and is flexible to explore the large dynamical range of astrophysical parameters. Parameterization of the cosmological signal has been done using physical parameters which are directly related to the IGM properties, from which we can infer the physics of the earliest sources. \citet{Pritchard_2010} proposed a turning point model, where the parameters of the model are the positions of the turning points (marked by points B, C, D in Fig.~\ref{fig:global}) - these are positions in redshift or frequency, and in brightness temperature. The signal between these turning points is modelled as a cubic spline. This parameterization is very flexible and can describe a wide range of 21cm signals, but the turning point positions require further interpretation in order to relate them to the physics of the first sources \citep{Mirocha_2013}. The $\tanh$ parameterization \citep{Harker_2015,Mirocha_2015}, uses a set of $\tanh$ functions to model the Global signal. In \citet{Bernardi_2015,Bernardi_2016}, the absorption feature is modelled as a Gaussian.  \\
In this paper, we have used $\tanh$ parameterization to simulate the Global 21cm signal as proposed in \citet{Mirocha_2015}. This method employs modelling the $Ly\alpha$ background, IGM temperature and  ionization fraction as simple $\tanh$ functions \citep{Harker_2016}. The $Ly\alpha$ background determines the strength of Wouthuysen-Field coupling, $T$ the temperature of the IGM, and the $\overline{X}$ is the ionized fraction of hydrogen. We let each quantity evolve as a $\tanh$ function \citep{Mirocha_2015} given by:
\begin{equation}
A(z)=\frac{A_{\mathrm{ref}}}{2}\lbrace 1+tanh[(z_{0}-z)/\Delta z] \rbrace,
\label{eq:tanh}
\end{equation}
where $A(z)$ represents the parameters $J_{\alpha}(z)$, $T(z)$ and $\overline{X}_{i}(z)$. The free parameters of the $\tanh$ model are the step height, $A_{ref}$, pivot redshift, $z_{0}$ and a width or duration, $\Delta z$. $J_{\alpha}(z)$ represents the $Ly\alpha$ background which determines the strength of the Wouthuysen-Field coupling. $T(z)$ represents the temperature of the IGM and $\overline{X}_{i}$ is the ionization fraction. These parameters are directly linked with IGM properties but not to the source properties. So, in some sense this is an intermediate between the physical modeling and phenomenological models, like the spline or Gaussian models. \\
\begin{figure}
    \centering
    \includegraphics[width=\columnwidth]{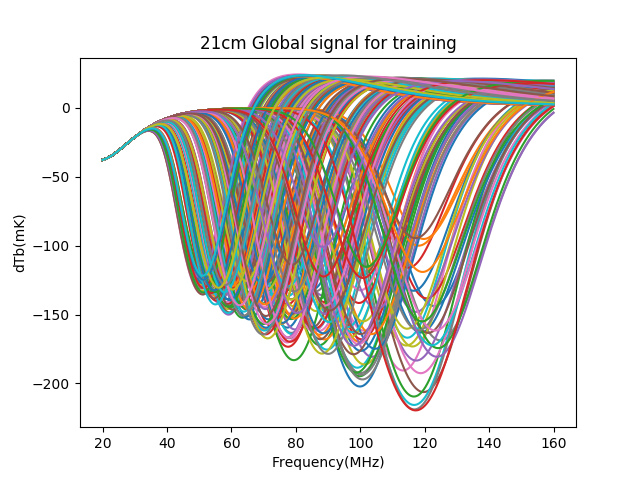}
    \caption{The set of 21cm Global signals generated to construct the training datasets, by varying the signal parameters.}
    \label{fig:GlobalSet}
\end{figure}
In this paper, we have used three $\tanh$ functions representing the parameters $J_{\alpha}(z)$, $T(z)$ and $\overline{X}_{i}(z)$ which in turn are functions of three parameters each as listed below:
\begin{equation}
\centering
\begin{split}
 J(z)=\frac{J_{\mathrm{ref}}}{2}\lbrace 1+tanh[(J_{z0}-z)/J_{dz}] \rbrace   \\
\overline{X_i}(z)=\frac{X_{\mathrm{ref}}}{2}\lbrace 1+tanh[(X_{z0}-z)/X_{dz}] \rbrace   \\
T(z)=\frac{T_{\mathrm{ref}}}{2}\lbrace 1+tanh[(T_{z0}-z)/T_{dz}] \rbrace \\ 
\end{split}
\end{equation}
The normalization $A_{\mathrm{ref}}$, (Eqn.~\ref{eq:tanh}) of the $\tanh$ function corresponding to the $Ly\alpha$ flux is $J_{\mathrm{ref}}$, in the units of $10^{-21}\ \mathrm{erg~s^{-1}~cm^{-2}~Hz^{-1}~sr^{-1}}$. The redshift interval $(\Delta z)$ and the central redshift $(z_{0})$ over which the $Ly\alpha$ background turns on, are represented by $J_{dz}$ and $J_{z_{0}}$ respectively. For X-ray heating, $T(z)$ is in units of $K$, $\Delta z$ and $z_{0}$ are denoted by $T_{dz}$ and $T_{z_{0}}$. The step height corresponding to $T(z)$, i.e., $T_{\mathrm{ref}}$ is fixed at 1000 K. As the signal saturates at low redshifts, so the precise height of the step is not important. The step height corresponding to the ionization fraction, $\overline{X}_{i}$ is fixed to unity, since it represents a fraction.
So, we have 7 signal parameters to construct the signal using the $\tanh$ parameterization.
We compute the coupling coefficients using \textit{ARES}, and plug in the values of the parameters into Eqn.~\ref{eq:global} to obtain the simulated Global 21cm signal. 
Hence, our parametrized Global 21cm signal depends on total 7 parameters (with 2 parameters, $X_{\mathrm{ref}}=1.0, T_{ref}=1e3$, fixed). We take the inferred values of these parameters from \citet{Harker_2015}$:J_{\mathrm{ref}}=11.69,  J{z_{0}}=18.54, X{z_{0}}=8.68, T_{z_{0}}=9.77,
J_{dz}=3.31, T_{dz}=2.82, X_{dz}=2.83 $. Each of these parameters are varied by $\pm50\%$ to generate our training sets. The parameter range explored here is sufficient to represent a wide range of shapes of the Global signal. A typical set of such signals generated for this work is shown in Fig:~\ref{fig:GlobalSet}.
The reason why we choose the $\tanh$ model is because, it can mimic the shape of the Global 21cm signal very well, and can be related to the physical properties of the IGM to a great extent. The $\tanh$ parameters are directly related to the IGM properties, but cannot give us information about the source properties directly. Thus, it can be considered to be somewhere between the `turning point' model, which is completely phenomenological and other entirely physical models. 
\begin{figure}
\includegraphics[width=\columnwidth]{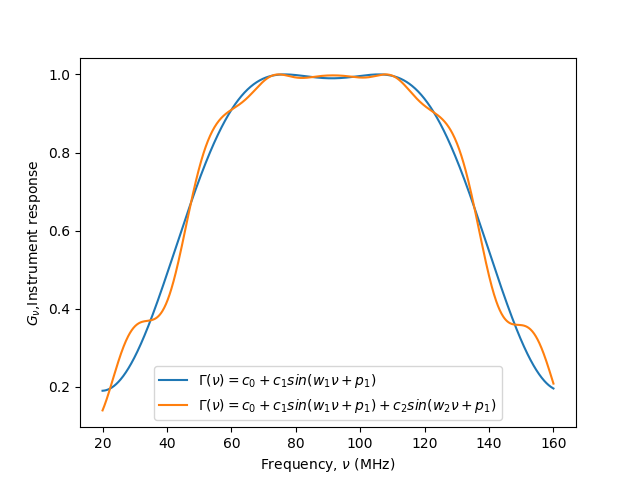}
\caption{Here we have plotted the instrument response represented by $G(\nu)=1-|\Gamma|^{2}$, where $\Gamma(\nu)$is the antenna reflection coefficient for a simple instrument and a moderate instrument. The parameters were chosen so that $G(\nu)$ would resemble a bandshape.}
\label{fig:Instruresponse}
\end{figure}

\section{Foregrounds}\label{foregrounds}

The faint Global 21cm signal from Cosmic Dawn/Epoch of Reionization has to be detected in presence of bright foreground sources in the sky and instrumental/atmospheric corruptions. Radio emission from our Galaxy and other extragalactic sources are many orders of magnitude brighter than the cosmological signal of interest. The expected Global 21cm signal is about $10^{-4}$ times weaker than the foreground emission. The bright foregrounds, Radio Frequency Interference (RFI), instrumental calibration errors poise significant challenges to Global 21cm experiments. Hence, several sophisticated simulations are necessary to understand the effect of these corruption terms on the possible signal extraction method. That makes it critical to have an accurate model for the foregrounds at these radio frequencies. 

Following \citet{Pritchard_2010,Bernardi_2015}, the foreground spectrum can be modelled as a polynomial in log($\nu$)-log(T). While \citet{Harker_2015} showed that use of 3rd or 4th order polynomial is sufficient to represent the sky spectrum, \citet{Bernardi_2015} showed that a 7th order polynomial is necessary when incorporating the chromatic primary beam of the antenna.
In our case, we restrict our foreground model to 3rd order polynomial in  $\ln(T)$-$\ln(\nu)$ \citep{Harker_2015}, representing diffuse foregrounds:
\begin{equation}
\ln{~T_{\mathrm{FG}}}=\sum_{i=0}^{n}~a_{i}[\ln{(\nu/\nu_{0})}]^{i},
\label{eq:fg}
\end{equation}
where, $\nu_{0}$ is taken to be $80$MHz, which is an arbitrary reference frequency and ${T_{0}, a_{1}, a_{2},...a_{n}}$ are the parameters of the model, where $a_0=\ln{T_0}$, as the zeroeth order coefficient has units of temperature. Increasing $n$ implies more complex, less smooth foregrounds. Here we haven taken the values of $ \{T_{0},~a1,~a2,~a3\}=~\{ 2039.611,-2.420 96,-0.080 62,0.028 98\}$ \citep{Harker_2015, Costa_2008}. For the simulations, we varied them as $\{15\%,10\%,1\%,1\%\}$ respectively. As the foregrounds are spectrally smooth as compared to the cosmological signal, this feature is taken advantage of in signal extraction.\\

\section{The Instrument}\label{instrument}
 In this section, we introduce the model for the instrument or the radio telescope using which the detection of the cosmological signal can be attempted. Although it appears to be a very simple case to observe with a total power radiometer, it is actually quite complex to expect the stability of the instrument over such a large bandwidth of frequencies and over large hours of observing time \citep{Burns_2012,Burns_2017}. The calibration of such an instrument becomes quite a challenge as any residual calibration error might mimic the cosmological signal. Hence, understanding of the instrument stability and robustness of our proposed signal extraction technique with instrumental calibration error is critical. 

In \citet{Harker_2012, Harker_2015} or other related works, the analyses mostly consider a perfect instrument, with no residual calibration error or impact on the foreground removal. Frequency and angular dependence of real antenna response complicates the foreground modelling, commonly based on the assumption of spectral smoothness. The coupling between the foregrounds and  antenna gain pattern generates spectral structure in the foregrounds that is required to be represented by additional higher order terms in the foreground. 

Here, we have followed a model of the instrument as described for DARE (Dark Ages Radio explorer) \citep{Bradley_2012, Burns_2017}, where a set of sinusoids is fitted to the modeled antenna reflection coefficient, $\Gamma(\nu)$; 
\begin{equation}
\begin{split}
\Gamma(\nu)=c_{0}+c_{1}~\sin(\omega_1\nu+p_{1})+c_{2}~\sin(\omega_2\nu+p_{2})\\
+c_{3}~\sin(\omega_3\nu+p_{3})+c_{4}~\sin(\omega_4\nu+p_{4})
\end{split}
\end{equation}
And the instrument response or the gain function is represented by, 
\begin{equation}
G(\nu)=1-|\Gamma(\nu)|^{2}
\label{eq:instru-response}
\end{equation} 
We have considered two cases for the instrument: (a) simple and (b) moderate. We call the (a) \textit{simple} instrument case, where $\Gamma$ is represented by a single sinusoid only. We chose the parameters of $\Gamma$ such that, the response mimics a bandpass shape. We have chosen, $c_{0}=0.4; c_{1}=0.5; w_{1}=0.044; p_{1}=0.7$ [See Fig.~\ref{fig:Instruresponse}]. 
\begin{equation}
\Gamma(\nu)=c_{0}~+~c_{1}~\sin(\omega_{1}\nu + p_{1})
\label{eq:instru1sine}
\end{equation}

We call the other one as (b) \textit{moderate} instrument, where we have assumed that the instrument response can be represented as a sum of two sinusoids-one with a large amplitude and a low frequency and another with a very small amplitude and a high frequency .
\begin{equation}
\Gamma(\nu)=c_{0}~+~c_{1}~\sin(\omega_{1}\nu+p_{1})+c_{2}~\sin(\omega_{2}\nu)
\label{eq:instru2sine}
\end{equation}
The chosen value of the parameters for this case were: $ c_{0}=0.4, c_{1}=0.5, \omega_{1}=0.044, p_{1}=0.7, c_{2} =-0.05, \omega_{2}=0.25 $ [See Fig.~\ref{fig:Instruresponse}].
We have varied each of the instrument parameters by $\sim5\%$ for all our simulations in this work.
\section{Artificial Neural Networks}\label{ANN-intro}
\subsection{Introduction: Basic architecture}
\acrlong{ann} are information processing systems, whose performance characteristics are inspired by the structure and functioning of the human brain and the nervous system. ANNs simulate the learning process from examples mimicking the way humans tend to learn. There can be different kinds of supervised and unsupervised learning algorithms, which can be implemented to suitably teach and train a network as per requirement.\\ 
\begin{figure}
\includegraphics[width=3.25in]{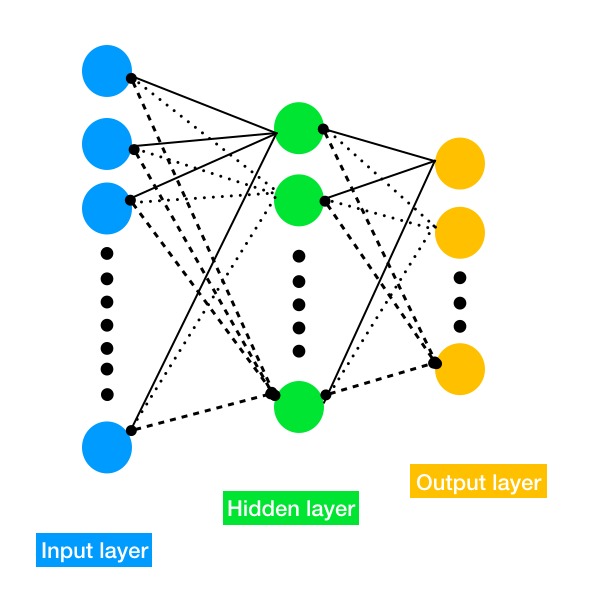}
\caption{Schematic representation of a typical neural network. Each coloured circle represents a neuron, which form the basic units of an ANN. The lines represent the connections from neurons of one layer to the next. The blue, green and the orange layers represent the input, hidden and the output layers of the network. It should be noted that there can be more than one hidden layer in an ANN. }
\label{fig:ANNstructure}
\end{figure}
The basic building blocks of the \acrshort{ann}s are the \textit{neurons}. The most simple ANN usually is a three-layer network comprising of the input layer, a hidden layer and the output layer [see Fig.~\ref{fig:ANNstructure}]. The number of neurons in the hidden layers decide how \textit{wide} the network is and the number of hidden layers decide how \textit{deep} the network is. The final structure of the network is decided after taking into consideration several factors, which will be dealt in detail in the following sections.
\begin{figure*}
\includegraphics[width=0.8\paperwidth]{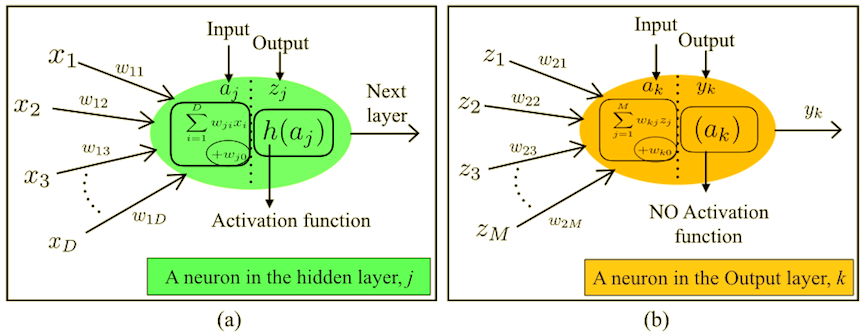}
\caption{Typical neuron structures explaining what exactly happens inside the neurons of the hidden layer (left) and the output layer (right). Each neuron receives the input from all the neurons in the previous layer, along with their respective weights and biases.} 
\label{fig:neurons}
\end{figure*}
As briefly outlined in \citet{choudhury_2017}, in this section, we introduce in detail how a neural network works.
ANNs construct functions, which associate the input with the output data. The basic neural network model can be described as a series of functional transformations. The \textit{input layer}, is the first layer of the network. We consider that there are D inputs, $(x_{1},x_{2},...,x_{D})$. Each neuron in the input layer (which will be denoted by an index $i$) is connected to each neuron in the next layer, which is the hidden layer (denoted by an index $j$). A weight $w_{ji}$ and a bias $w_{j0}$ are associated with each connection. The input to a single neuron in the hidden layer is a linear combination of the input neurons with their respective weights and biases, and is written as:
\begin{equation}
a_{j}^\mathrm{H}=\sum_{i=1}^{D} w_{ji}^{\mathrm{I\rightarrow H}}~x_{i}+w_{j0}^{\mathrm{I\rightarrow H}}
\end{equation}
where, $w_{ji}~'s$ are the weights and $w_{j0}~'s$ are the biases associated with each connection between the neurons of the input and hidden layer respectively, denoted by the superscipt $\mathrm{I\rightarrow H}$. These $a_{j}'s$, are known as \textit{activation} [Fig.~\ref{fig:neurons}] and the superscript $\mathrm{H}$ represents a hidden layer neuron. In the hidden layer, each of the activations is transformed using a non-linear activation function $h$, such that,
\begin{equation}
z_{j}=h(a_{j}^\mathrm{H})
\end{equation}

These quantities $z_{j}'s$ correspond to the outputs of each neuron in the hidden layer. These activation functions are very interesting and can be chosen to be non-linear functions, such as the logistic sigmoid function or $\tanh$ function. These functions are smooth and differentiable, and it saturates and returns a constant output when the absolute value of the input is sufficiently large. It is because of the activation function that a non-linearity is introduced in the network and the trained ANN performs appropriately. Otherwise, the entire algorithm would just be solving some linear set of functional transformations. 
These values $(z_{j}'s)$ are again linearly combined along with the corresponding weights and biases associated with the respective connections to give the inputs to each neuron in the output layer, denoted by the index $k$. These are also called the output unit activations, $a_{k}^\mathrm{O}$, given by: 
\begin{equation}
a_{k}^\mathrm{O}=\sum_{j=1}^{M} w^{\mathrm{H\rightarrow O}}_{kj}~z_{j}+w^{\mathrm{H\rightarrow O}}_{k0}
\end{equation}
here, $w_{kj}$'s and the $w_{k0}$'s are the weights and biases respectively, connecting the hidden and the output layer, denoted by the superscript $\mathrm{H\rightarrow O}$.
The output unit activations are then transformed using an appropriate activation function, to give a a set of network outputs $y_{k}$. [See Fig.~\ref{fig:neurons}]

\begin{equation}
y_{k}=h(a_{k}^{\mathrm{O}})
\end{equation}

These $y_{k}'s$ are the end outputs of the \textit{feed-forward process} and $h$ is the activation function. It must be noted that the activation function is usually not applied to the output layer, or in most cases it is a linear function. So far, we have given a input vector X, of dimension D and obtained an output vector Y, of our desired dimension, chosen as per requirement. 
\subsection{The Training Process}
\acrshort{ann}s are viewed as a general class of parametric non-linear functions from an input variables vector $X$ to a vector $Y$ of output variables. The parameters in question are the respective weights and biases. 
These parameters are determined usually by minimizing a sum-of-squares error function. We denote this error function by $E(w)$. Error functions for a set of independent observations comprise of sum of terms, one for each data point,
\begin{equation}
E(w)=\sum^{N}_{n=1} E_{n}(w)
\end{equation}
This is the total \textit{error function} or the total \textit{cost function}.
The goal is to find a suitable weight vector $W$ which minimizes the chosen function $E(w)$. A very small step in the weight space from $w$ to $w+\delta w$, changes the error function by a quantity, $\delta E\simeq \delta w^{T}\nabla E(w)$, where the vector $\nabla E(w)$ points in the direction of the greatest rate of increase in the error function. As the error $E(w)$ is a smooth continuous function of w, its smallest value will occur at a point in the weight space such that the gradient of the error function vanishes, so that $\nabla E(w)=0$. Else, we could make a small step in the direction of $-\nabla E(w)$ and further reduce the error. The error function typically has a nonlinear dependence on the weights and bias parameters, so there will be many points where the gradient vanishes or becomes very small. As we cannot directly find the solution to the equation $\nabla E(w)=0$, numerical procedures are used to find the solution. Some initial value, $w^{(0)}$ is chosen for the weight vector and then we move through the weight space in succession of small steps till error function is minimized. This optimizing is done using a stochastic gradient decent, or \textit{adam} \citep{Kingma_2014}. Adam deals with outliers very well and has been proven to be a better optimizer than the standard stochastic gradient descent, which is discussed in detail in \citet{Kingma_2014}. So, the outputs obtained are compared with the inputs with which the network was fed and an error function is computed. We find out an optimum set of weights and biases, that ensures that the output produced by the network is sufficiently close to the desired output values. 

\subsection{Back-propagation}
The error function $E$ is a function of the weights. The error computed at the end of the output layer, at each node (output of the neurons) is the difference between the target and the actual value obtained. The node error can be defined as:
\begin{equation}
e_{k}=t_{k}-o_{k}
\end{equation}
where, $t_{k}$ is the target and $o_{k}$ is the output obtained at node k, in the output layer. 

\begin{figure}
\includegraphics[width=\columnwidth]{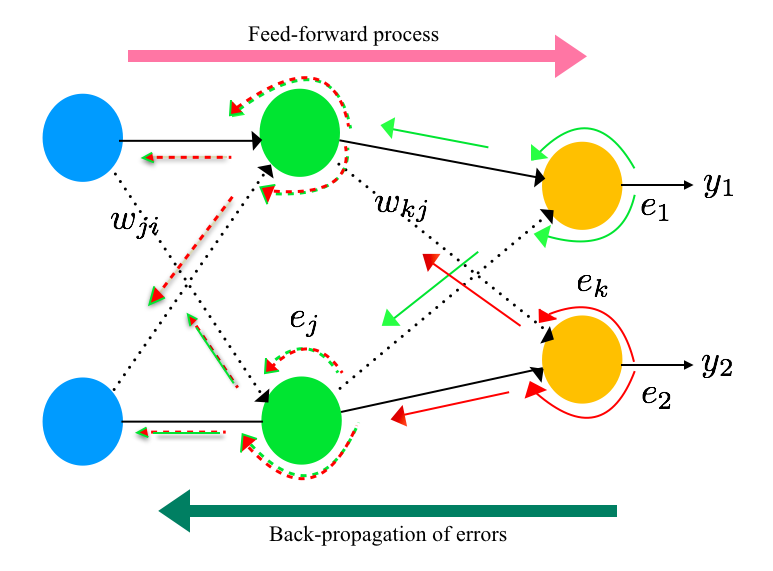}
\caption{The back propagation algorithm schematically explained. The coloured arrows show how the error is propagated backwards, while the solid black arrows demonstrate the feed forward process.}
\label{fig:bpp}
\end{figure}

The \textit{back-propagation algorithm}, is referred to as \textit{backprop} or BPP. The errors are `back-propagated' to adjust the weights such that the error function is minimized. The total error is proportionately propagated backwards. The higher weights are considered to be more responsible for the errors, thus are proportionately split to the connected links [See Fig. ~\ref{fig:bpp}]. 

The errors associated with the internal links are not so straight-forward. The output layer errors are split in proportion to the connection weights and recombine the bits of error at each node in the hidden layer.
The slope of the error function is calculated, so that the weights can be accordingly adjusted.
\begin{equation}
\frac{\partial E}{\partial w_{kj}^\mathrm{H\rightarrow O}}=-(e_{k})h\left(\sum_{j}w_{kj}^\mathrm{H\rightarrow O}.o_{j}\right)\cdot \left[1-h\left(\sum_{j}{w}_{kj}^\mathrm{H\rightarrow O}.{o}_{j}\right)\right].{o}_{j}
\end{equation}
Here, $h$ is the activation function and the superscripts represents the connections to which the weights are associated. Similarly, the weights $w_{ji}$ are adjusted as,
\begin{equation}
\frac{\partial E}{\partial w_{ji}^\mathrm{I\rightarrow H}}=-(e_{j})h\left(\sum_{i}w_{ji}^\mathrm{I\rightarrow H}.o_{i}\right)\cdot
\left[1-h\left(\sum_{i}w_{ji}^\mathrm{I\rightarrow H}.o_{i}\right)\right].o_{i}
\end{equation} 
Here, $e_{j}$ is the recombined back-propagated error from the hidden nodes [See Fig.~\ref{fig:bpp}].

The weights are updated using the above relations. It is to be noted that the weights are changed in a direction opposite to the gradient. A quantity $\alpha$ is introduced which is called the learning rate.
\begin{equation}
\mathrm{new~w_{kj}^\mathrm{H\rightarrow O}=old~w_{kj}^\mathrm{H\rightarrow O}-\alpha.\frac{\partial E}{\partial w_{kj}^\mathrm{H\rightarrow O}}}
\end{equation}

After each update the gradient is re-evaluated for the new weight vector and this process is repeated. An important point to be noted is that the error function is defined with respect to a training set, and the entire training set is processed at each step, in order to compute the gradient. Such techniques which use the entire dataset at once are called batch methods. At each step the weight vector is moved in the direction of the greatest rate of decrease of the error function. This update is repeated by cycling through the data either in sequence or by selecting points at random with replacement. 



\section{Artificial Neural Networks - Another Technique to extract the Global 21cm Signal}\label{ANN-21cmGS}
The technique of ANN can be implemented to extract a faint cosmological signal from a significantly bright, dominant foreground. The most important requirement for a good model for predicting the parameters is a rich training dataset, which would contain several possible realizations of the Global signal, along with foreground added to it. We can build a good network by training, testing and validating it, the process has been described in detail in the above section. Here we have built an ANN which would estimate the signal and the foreground parameters. In the following sections we will describe in detail how we have formulated the problem and built the network.\\
\\
\\
\\
\section{Simulation}\label{simulation of21cmGS}
\subsection{Constructing the training dataset}
\begin{figure*}
\centering
\includegraphics[width=6.5in]{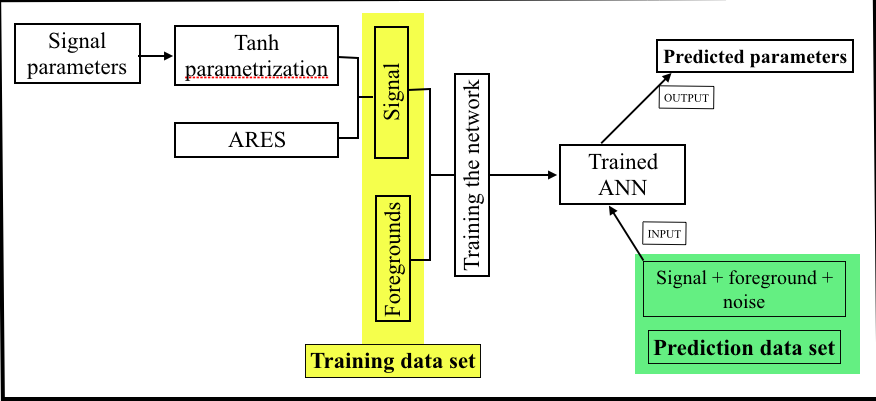}
\caption{ This shows how we have constructed the training datasets and built the network. Once the network is ready, we have fed the network with the prediction datasets, unknown to the network.}
\label{fig:Blockdiagram}
\end{figure*}

Our training dataset consists of 270 samples. Each sample in this training dataset is a total signal $(T_{tot}$), which contains one realization of the cosmological signal that has been generated using the $\tanh$ parameterization (\textsection\ref{21cmGS simulation}) $T_{21}$, along with the foreground, which is generated by the log-polynomial model (\textsection\ref{foregrounds}) $T_{FG}$. This constitutes the training dataset, in the simplest case. 
\begin{equation}
T_{\mathrm{train}}(\nu)=T_{21}(\nu)+T_{\mathrm{FG}}(\nu)
\end{equation}
In the cases where we add the instrument response $G(\nu)$, the training dataset is constructed as:
\begin{equation}
    T_{\mathrm{train}}(\nu)=(T_{21}(\nu)+T_{\mathrm{FG}}(\nu))\cdot G(\nu)
\end{equation}
A block diagram illustrating the entire process is shown in Fig.~\ref{fig:Blockdiagram}.

\begin{figure}
\includegraphics[width=\columnwidth]{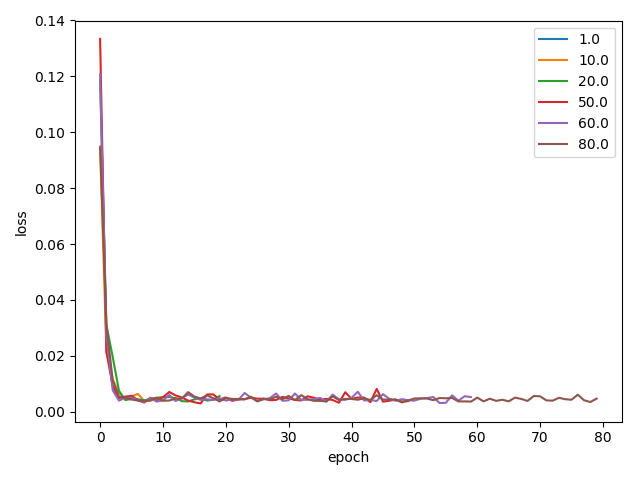}
\caption{The RMSE calculated for the different parameters were plotted for different networks with varying number of epochs. One \textit{epoch} is one complete pass of all the training samples, while an iteration is the total number of passes. Say, if we have 500 training samples, and a batch size of 100, then it will take 5 iterations to complete one epoch. We can see that for number of epochs greater than 50, all the parameters have a sufficiently low RMSE, which almost saturates. }
\label{fig:iters}
\end{figure}

\begin{figure}
\includegraphics[width=\columnwidth]{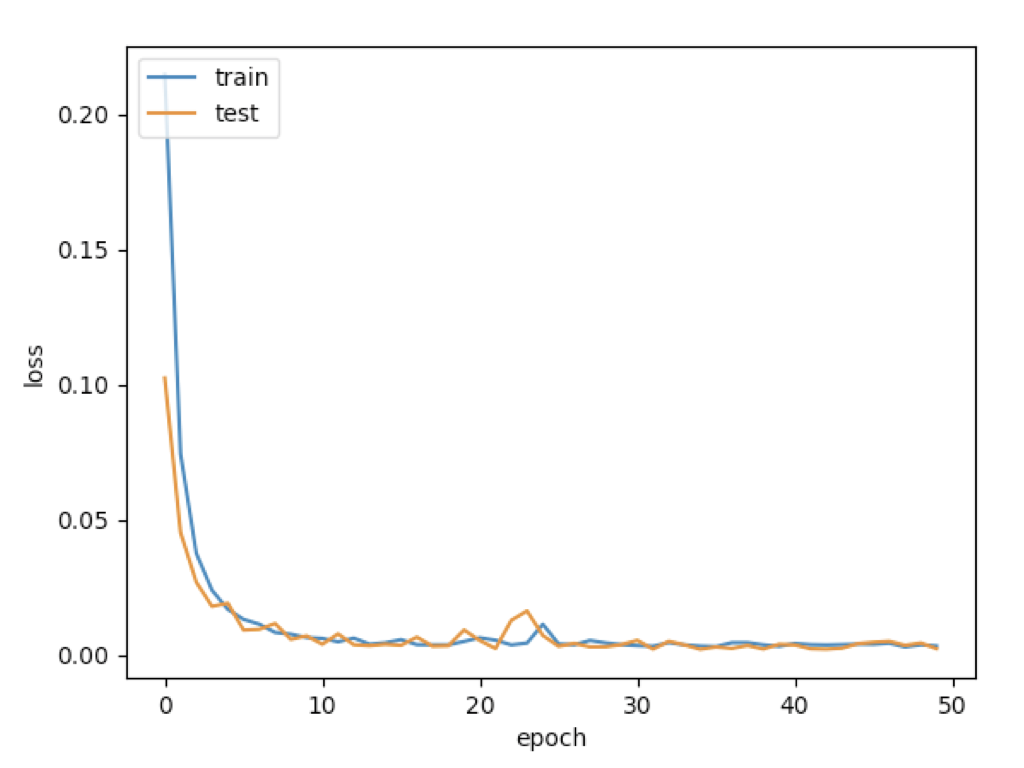}
\caption{ This is the model loss function for the perfect instrument case. We observe that for this case, the model loss saturates after around 50 epochs.}
\label{fig:modelloss}
\end{figure}

We use a feed-forward network, with two hidden layers. We have 1024 neurons in the input layer, corresponding to the $1024$ frequency channels we choose to work with, for a bandwidth of $20-160$ MHz, and 18 neurons in the first hidden layer, which is connected to the next hidden layer with 14 neurons. The choice of the number of neurons in the hidden layer is determined after checking how the error is evolving. We begin with an initial guess, and vary the number of neurons in the hidden layers and the number of hidden layers, and note the numbers around which the RMSE is the least. We then decide upon the final structure of the network. Then, we vary the number of iterations to look at how the RMSE of the predicted parameters would change (see Fig:~\ref{fig:iters}). We also look at the model loss, for those number of neurons, which should ideally decrease with increasing number of iterations and saturate after a certain point (see Fig.~\ref{fig:modelloss}). Taking all these into consideration, we decide upon a number. The hidden layer neurons are activated by a sigmoid function. 
The number of neurons in the output layer is easy to choose. We have 11 output neurons corresponding to each parameter we are interested to find out, i.e, 7 signal parameters and 4 foreground parameters. Our main aim is to extract the signal parameters and reconstruct the 21cm signal.
For each case that we have studied, we trained and used different networks to  predict the parameters.
We have used packages {\sc scikitlearn} \citep{Pedregosa_2011} and {\sc keras} to build simple neural networks. We construct a simple neural network with two Dense-layers, using the \textit{Sequential} model available in \textit{keras}.   

\subsection{Training, validation and testing the network}
The training process is an iterative procedure that begins by collecting the data and preprocessing it to make training more efficient. This stage involves normalizing the data, and dividing it into training, validation and testing sets. In this type of network, the information flow is unidirectional that takes place with the help of an activation function between each layer. In our network, the data is normalized using min-max normalization. The entire dataset is split into 2 parts: training and the testing set, in a 7:3 ratio. The training set is further split into chunks of validation sets, iteratively, during the training process. Once the training is complete, the network is tested on the testing set. \\
The network uses back-propagation [see Fig.~\ref{fig:bpp}] as the learning technique. We choose the mean-squared error as the error-function. After repeating this process for a sufficiently large number of training cycles, the network will usually converge to some state where the error function is the minimum. In this case, one would say that the network has learned a certain target function. To adjust weights properly, we use the \textit{`adam'} (Adaptive Moment Estimation) optimizer for non-linear optimization. The main advantage of choosing \textit{adam} over the commonly used \textit{stochastic gradient descent} is that, it uses the combined averages of previous gradients at different moments to give it more expressive power to better update the parameters, in a more adaptive manner. \\
After training the network, we analyse the performance of the network. This analysis may lead us to discover problems with the data, network architecture, or the training algorithm. The entire procedure is repeated until the network performance is satisfactory. As we run a particular model for training, at each step we get a validation score, in the form of validation accuracy and model loss. In our models we had obtained a validation accuracy of around 80 percent.  \\
\subsection{Constructing the prediction datasets}
The prediction set consists of samples which are unknown to the network and are used to test the robustness of our "trained network". In order to construct the prediction dataset, we use the models of 21cm signal and foreground, and add instrumental thermal noise to these datasets. We have produced a set of 90 prediction datasets following the equation:

\begin{equation}
    T_{\mathrm{pred}}(\nu)=T_{21}(\nu)+T_{\mathrm{FG}}(\nu)+ n(\nu)
\label{eq:pred}    
\end{equation}
In the cases where we have added the effect of the instrument, Eqn:~\ref{eq:pred} modifies as, 

\begin{equation}
    T_{\mathrm{pred}}(\nu)=(T_{21}(\nu)+T_{\mathrm{FG}}(\nu))\cdot G(\nu)~+ n(\nu)
\end{equation}
We use $T'_{sig}(\nu)$ to represent the prediction datasets in the following sections.
We introduce the instrumental thermal noise corresponding to a certain hours of observations ($N_{t}$) to the prediction dataset.\\
From the ideal radiometer equation, the noise, $n(\nu)$, in the observed spectrum can be written as:
\begin{equation}
n(\nu) \approx\frac{T_{\mathrm{sys}}(\nu)}{\sqrt{\Delta\nu\cdot\tau}}
\end{equation}
where, $\Delta\nu$ is the bandwidth of the observation and $\tau$ is the observation time. At these low radio frequencies, $T_{sys}$ is dominated by the brightness temperature due to the foregrounds. Hence, the above equation can be re-written as:
\begin{equation}
n(\nu)=\frac{T_{\mathrm{FG}}(\nu)}{\sqrt[]{\Delta\nu\cdot10^{6}\cdot3600\cdot N_{t}}}
\label{eq:thermal_noise}
\end{equation}
where, bandwidth $\Delta\nu$ is in MHz and $N_{t}$ is in hours. The factors $10^6$ and $3600$ are the respective conversion factors to the standard units.\\
Here, we are dealing with mock observations which are in turn simulated using some assumed instrument response and observational strategy. In future, similar network will be used for real observations to predict the redshifted Global 21cm signal. For real observations, the prediction data will be replaced by the actual data. The error estimation in case of real observations will then be computed from the error we get during training the network for `test' dataset. 

\section{Results}\label{results}
In this section, we will discuss results from simulations depicting different signal extraction scenarios: no instrumental effect, time independent instrumental effect and time dependant instrumental effect. 
\begin{figure}
\includegraphics[width=\columnwidth]{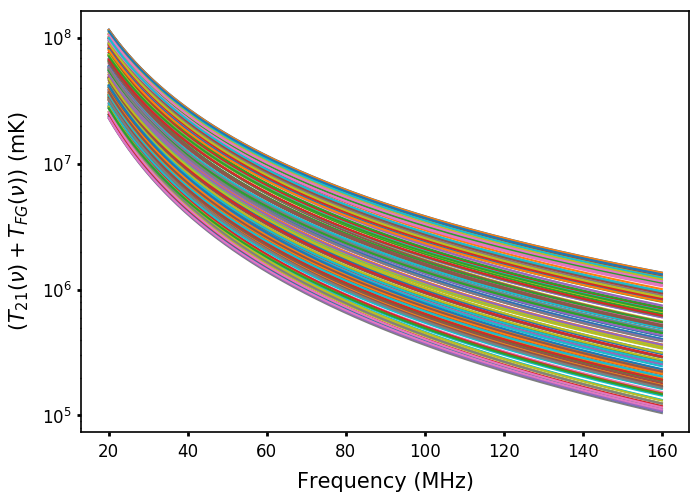}
\caption{Training dataset for the perfect instrument (Case 1). Note how the signal is entirely dominated by the foregrounds.}
\label{fig:perfecttraining}
\end{figure}

\begin{figure*}
\includegraphics[height=4.5in]{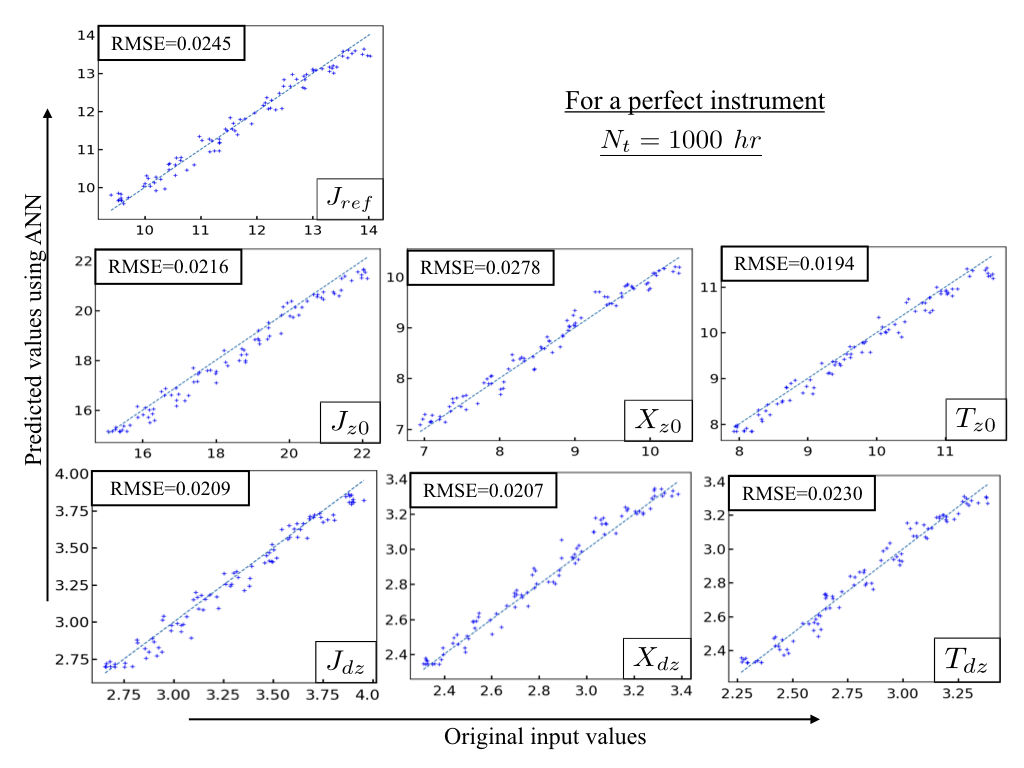}
\caption{Case 1-Perfect Instrument. The true values vs predicted values of the signal parameters are plotted. The dashed straight line in each plot represents the true values of the parameters. }
\label{fig:perfectrmse}
\end{figure*}

\begin{figure}
\includegraphics[width=\columnwidth]{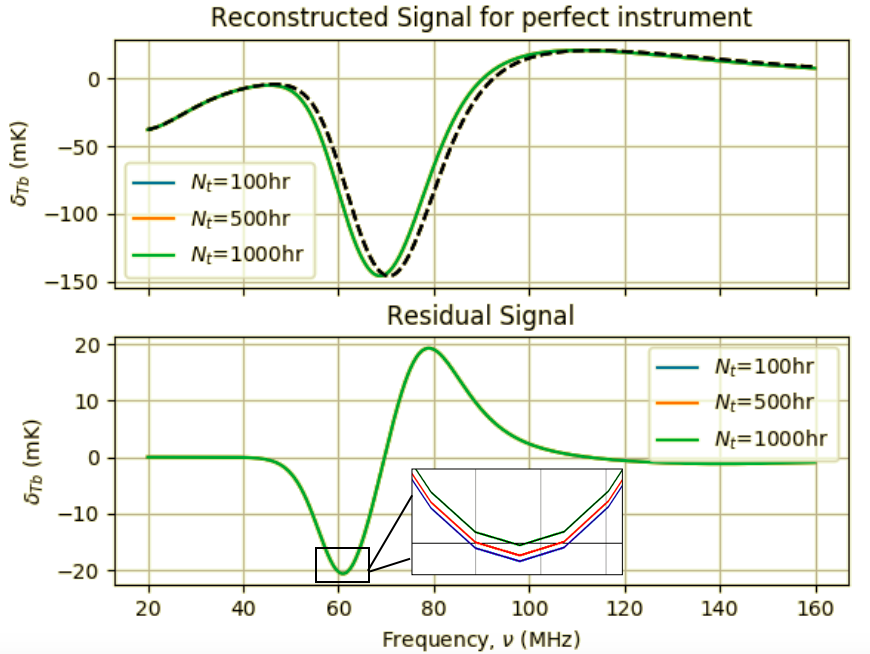}
\caption{Case 1: Reconstructed signal for the different $N_{t}$ assuming a perfect instrument (top panel). The residual signal, is plotted in the bottom panel,for increasing observation time, $N_{t}$. The zoomed section shows clearly how the residual decreases with increasing observation time.}
\label{fig:Subplotperfect}
\end{figure}

\subsection{Case 1: A Perfect Instrument}\label{case1_perfect}
For the simplest case, we take the 21cm signal along with the model foreground, assuming that the instrument is perfect. This is an ideal case where the instrument does not modify the signal in any way. The training dataset is constructed using the following relation:
\begin{equation}
T_{\mathrm{sig}}(\nu)=T_{21}(\nu)+T_{\mathrm{FG}}(\nu)
\end{equation}
where, $T_{21}$ is the Global 21cm signal and $T_{FG}$ is the model foreground (Eqn:~\ref{eq:fg}) as described in  \textsection\ref{foregrounds}. In Fig.~\ref{fig:perfecttraining} we show the training dataset, which is the total signal, $T_{sig}$ in mK units, including the cosmological signal and the foreground. The network is trained using this dataset and saved. We construct a prediction dataset containing 90 samples, using the following equation:
\begin{equation}
T_{\mathrm{sig}}'(\nu)=T_{21}(\nu)+T_{\mathrm{FG}}(\nu)+ n(\nu)
\end{equation}
where, $n(\nu)$ is given by Eqn:~\ref{eq:thermal_noise}. We use the saved network to predict the parameters from the prediction dataset. So we obtain 90 sets of predictions, for each sample in the prediction dataset.
Following \citet{Shimabukuro_2017}, we calculate a normalized RMSE for the predictions:
\begin{equation}
    \mathrm{RMSE}=\sqrt{\frac{1}{N_{\mathrm{pred}}}\sum_{i=1}^{N_{\mathrm{pred}}} \left(\frac{Y_{\mathrm{ori}}-Y_{\mathrm{pred}}}{Y_{\mathrm{ori}}}\right)^{2}}
\label{eq:RMSE}    
\end{equation}
where, $\mathrm{Y_{ori}}$ and $\mathrm{Y_{pred}}$ are the original and predicted values respectively, of one of the parameters, $Y\in[J_{ref},J_{z0},J_{dz},X_{z0},X_{dz},T_{z0},T_{dz}]$. $\mathrm{N_{pred}}$ denotes the total number of samples in the prediction dataset. The original and the predicted values of each of the parameters, along with their RMSE are plotted in Fig.~\ref{fig:perfectrmse}. Lower value of RMSE implies more accurate prediction of the parameters. When the instrument is considered to be perfect, we can see that all the parameters are predicted with RMSE $\approx~0.02$ [See Fig:~\ref{fig:perfectrmse}]. \\
In order to check the effect of the instrumental thermal noise on detection of Global 21cm signal, we have taken one sample 21cm Global signal and have varied the thermal noise, for observation times, $N_{t} \in (100, 500, 1000)$~hours [see Eqn:~\ref{eq:thermal_noise}]. We have computed the corresponding residuals, which is the difference between the input signal and the reconstructed signals, and plotted them in Fig.~\ref{fig:Subplotperfect}. The reconstructed signal appears to have overlapped for all the $N_{t}$. But in the zoomed section, it can be clearly seen that the residual is least for the signal with 1000 hours of observation (blue curve, see Fig.~\ref{fig:Subplotperfect}).

\subsection{Case 2: Imperfect instrument (fixed response)}
The instrument response modifies the signal considerably. By a fixed instrument response we mean that the exact behaviour of the instrument is known, it is very well calibrated, and it does not vary with changes in the surroundings, through the observation time.

\subsubsection{Simple instrument, fixed response}\label{case2a_fixed}
We have introduced the effect of an instrument response in \textsection \ref{instrument} (Eqn.~\ref{eq:instru-response}),
where the antenna reflection coefficient $(\Gamma)$ is represented by a set of sinusoidal harmonic basis functions. 
For a simple instrument, we consider only one sinusoidal term.
\begin{equation}
\Gamma(\nu)=c_{0}+c_{1}~\sin(\omega_{1}\nu)
\end{equation}
\begin{figure}
\includegraphics[width=\columnwidth]{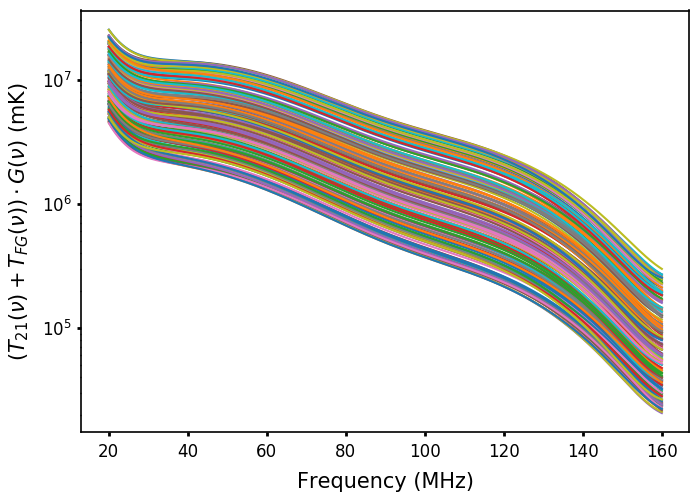}
\caption{Case 2A: Training dataset for a simple instrument with a fixed response whose $\Gamma$ is represented by one sinusoid. Note how the total signal is entirely dominated by the foregrounds and contaminated by the instrument effect.}
\label{fig:fixed1training}
\end{figure}
\begin{figure*}
\includegraphics[height=4.5in ]
{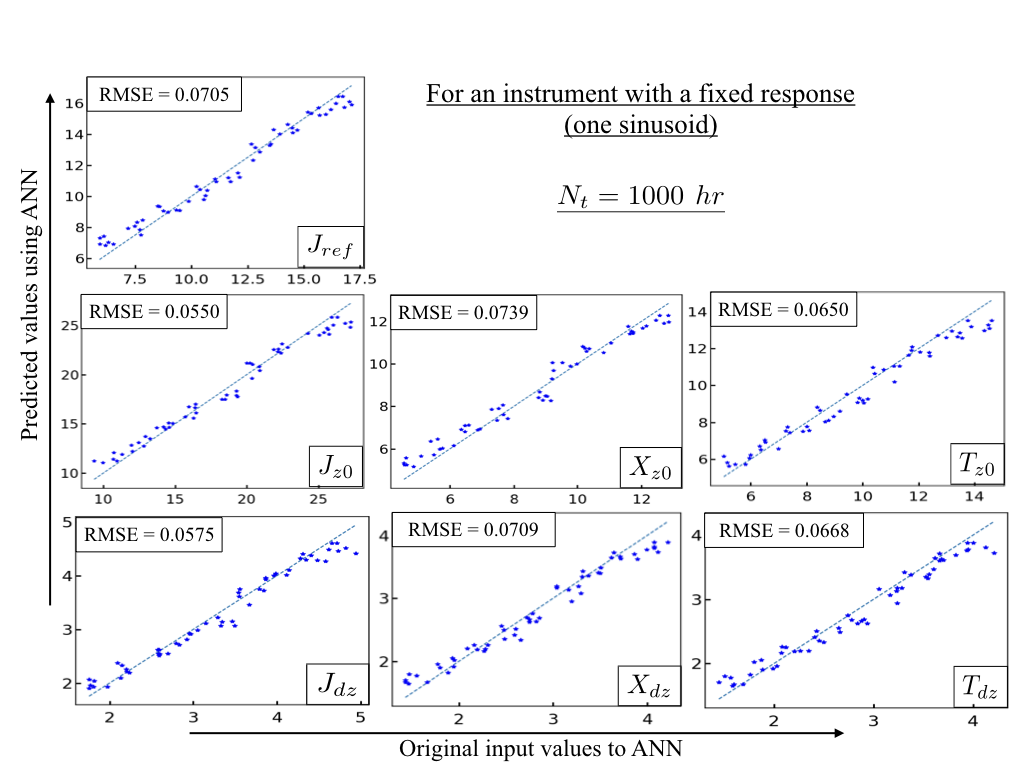}
\caption{Case 2A: Fixed response, simple instrument.The dashed straight} line in each plot represents the true values of the parameters, the dots represent the predicted values.
\label{fig:fixed1rmse}
\end{figure*}
\begin{figure}
\includegraphics[width=\columnwidth]{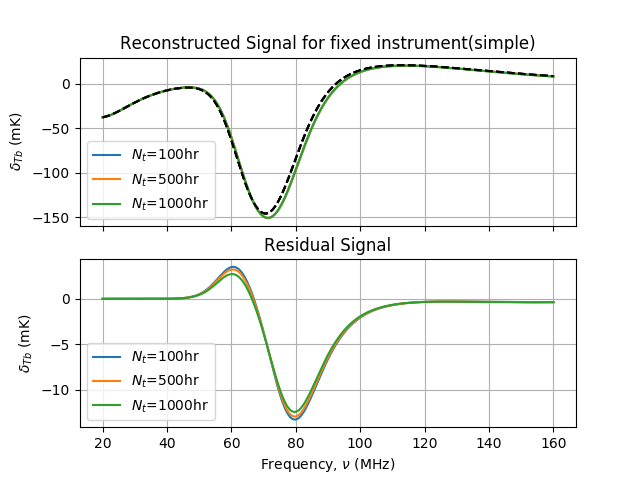}
\caption{Case 2A: The reconstructed signal (top panel) and the residual signal, with increasing observation time, $N_{t}$. The residual signal gives an idea of the quality of the reconstructed Global 21cm signal with the change in integration time.}
\label{fig:Subplot1fi}
\end{figure}
The training dataset is constructed following the relation: 
\begin{equation}
T_{\mathrm{sig}}(\nu)=(T_{21}(\nu)+T_{\mathrm{FG}}(\nu))\cdot G(\nu,c_{0},c_{1},\omega_{1})
\end{equation}
Here, the function $G(\nu)$ is the instrument response and $c_{0},c_{1},\omega_{1}$ are the parameters for the simple fixed instrument model.
The network is trained with this dataset (see Fig.~\ref{fig:fixed1training}) and saved.
The prediction dataset consists of a noisy data, which we feed into the network. The plots of the original versus the predicted values of the parameters, for the prediction dataset are shown in Fig.~\ref{fig:fixed1rmse}.
The RMSE for the parameters in this case is $\lesssim 7\%$, and has clearly deteriorated as compared to the perfect instrument case. \\
We then take a fixed signal and vary the total observation time, to see how the percentage error varies with increasing $N_{t}$. The reconstructed signals from the predicted parameters are shown in Fig.~\ref{fig:Subplot1fi}. From the plot of the residual signal, it is observed that the residual is least for the observation time, $N_{t}=1000~$hours shown by the green curve. 

\subsubsection{Moderate instrument(fixed response)}\label{case2b_fixed}
We train the network with a slightly more sophisticated instrument model, which we call a moderate instrument. We introduce an instrument response, whose $\Gamma$ consists of a combination of two sinusoids.
\begin{equation}
\Gamma(\nu)=c_{0}+c_{1}~\sin(\omega_{1}\nu) + c_{2}~\sin(\omega_{2}\nu)
\end{equation}
\begin{figure}
\includegraphics[width=\columnwidth]{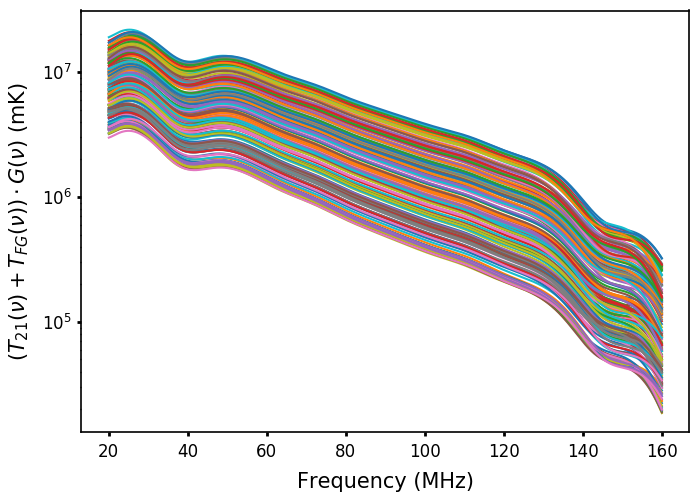}
\caption{Case 2B: Fixed response, moderate instrument ($\Gamma$ consists of 2 sinusoids): Training dataset for the network}
\label{fig:Fixed2training}
\end{figure}
The training dataset (see Fig.~\ref{fig:Fixed2training}) is constructed as:
\begin{equation}
\mathrm{T_{sig}(\nu)=(T_{21}(\nu)+T_{FG}(\nu))\cdot G(\nu,c_{0},c_{1},c_{2},\omega_{1},\omega_{2})}
\end{equation}
Here, $c_{0},c_{1},c_{2},\omega_{1},\omega_{2}$ are the parameters of the antenna reflection coefficient, and G has it's usual meaning. 
The noisy prediction dataset is then fed into the network.  
\begin{equation}
{\mathrm{T_{sig}'(\nu)=(T_{21}(\nu)+T_{FG}(\nu))\cdot G(\nu) + n(\nu)}}
\end{equation}
where, noise, n is given by Eqn.~\ref{eq:thermal_noise}.\\
\begin{figure*}
\includegraphics[height=4.5in ]{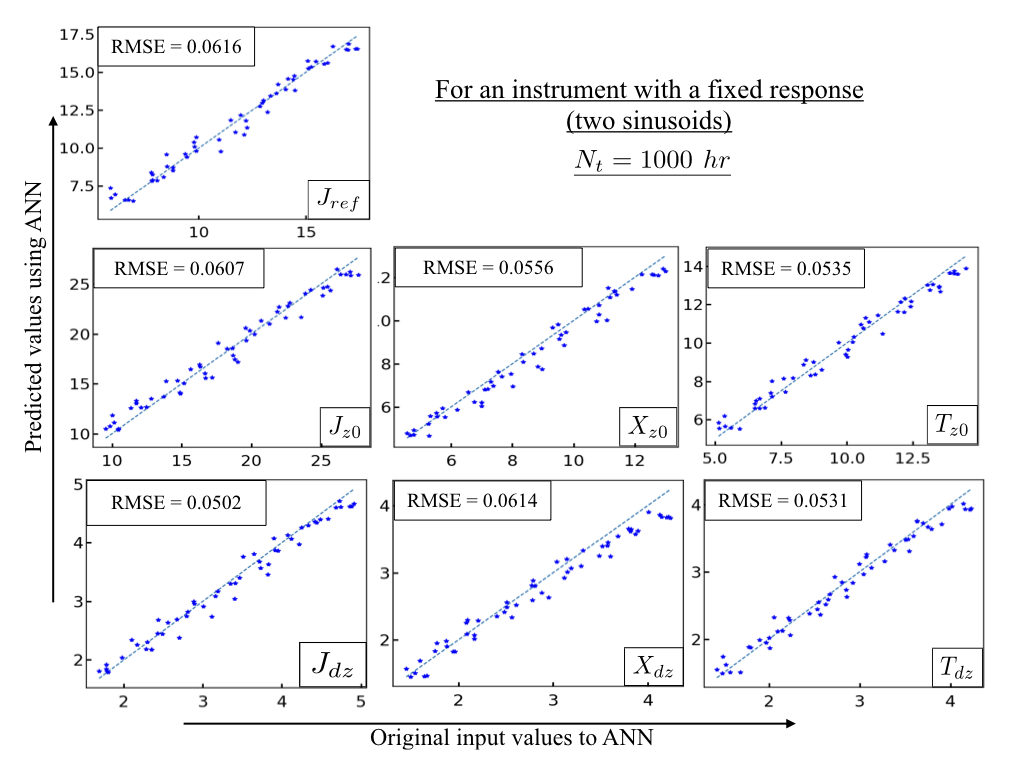}
\caption{Case 2B:Fixed response, moderate instrument.The true values vs predicted values of parameters are plotted. The dashed straight line in each plot represents the true values of the parameters.}
\label{fig:fixed2rmse}
\end{figure*}
The plots of the input versus the predicted values are shown in Fig.~\ref{fig:fixed2rmse}. 
\begin{figure}
\includegraphics[width=\columnwidth]{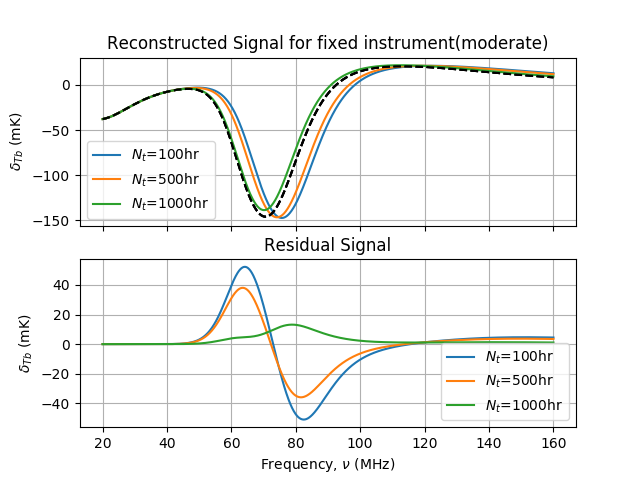}
\caption{Case 2B: The reconstructed signal (top panel) and the residual signal(bottom panel), with increasing observation time, $N_{t}$. The residual signal gives an idea of the quality of the reconstructed Global 21cm signal with the change in integration time.}
\label{fig:Subplot2fi}
\end{figure}
The reconstructed signals from the predicted parameters and the residuals are shown in Fig.~\ref{fig:Subplot2fi}. The RMSEs for the parameters in this case is $\lesssim 6\%$, and has also deteriorated as compared to the perfect instrument case.

\section{Summary and Discussions}\label{Discussions}
We have presented the detailed results obtained by using ANN to extract the cosmological Global signal, from a combined spectrum, which included bright dominant foregrounds, model instrumental effects, and thermal noise. We have considered two types of instrument models for this - a simple and a moderate instrument [see \textsection\ref{instrument}]. We have demonstrated how a simple ANN can easily handle up to 16 parameters (7 signal parameters, 4 foreground parameters, 5 instrument parameters, in the most complicated scenario which we have considered). The number of neurons in the hidden layer was optimized for each case for the best results. We summarize the calculated normalized RMSE, for each parameter, in each of the cases studied, in Table:~\ref{table:RMSE}.\\
\begin{table*}
\begin{tabular}{c|c|cc|cc} 
\hline
Parameters & Perfect & Fixed(simple) & Fixed(moderate) \\
 & Instrument & Instrument & Instrument \\
 & Case 1 & Case 2A & Case 2B \\
\hline
		$J_{\mathrm{ref}}$ & 0.0245 & 0.0705 & 0.0616  \\
 		$dz_{J}$ & 0.0209 & 0.0575 &  0.0502 \\
        $dz_{T}$ & 0.0230 & 0.0668 & 0.0531 \\
        $dz_{X}$ & 0.0207 & 0.0709 & 0.0614 \\
        $z0_{J}$ & 0.0216 & 0.0550 & 0.0607 \\
        $z0_{T}$ & 0.0194 & 0.0650 & 0.0535 \\
        $z0_{X}$ & 0.0278 & 0.0739 & 0.0556 \\
        
\hline
\end{tabular}
\caption{The RMSE values for the signal parameters, computed for the prediction dataset for each of the cases studied, are summarized here. Considering the perfect instrument case to be the reference, we can see how the errors increase in the cases where the instrument effect is considered.}
\label{table:RMSE}
\end{table*}
If we look at the RMSE's in Table: \ref{table:RMSE}, we see the effect of increasing the complexity of the input data on the accuracy levels. If the network is trained sufficiently well, then the accuracy levels continue to be high.  All errors are less than $10\%$, the worst prediction being  around $\sim 8\%$. The accuracy levels obtained are between $\sim 92-98\%$ for the signal parameters. The RMSEs for the cases which includes the fixed response $(2A~\&~2B)$, for the simple and moderate instruments, are not very different. This is because we are considering simple instrument models, where the moderate instrument model just has an additional sinusoidal term. However, if we compare to the perfect instrument case, we can see that the contamination affects the errors in predictions.
\\
To the best of our knowledge, most authors have used MCMC, nested sampling or similar methods to sample the parameter space. We cannot directly compare the speed of ANN with an efficient MCMC approach, as the methodologies are different. But, we bypass the task of computing the likelihood function, a large number of times, to arrive at the inferred values of the parameters, while using ANN. So, when dealing with a higher dimensional parameter space, ANN is faster and computationally more efficient. In the case of neural networks, we can think of the training sets as a more realistic equivalent to the prior in MCMC. \\
In \citet{Harker_2016}, Gaussian priors for the foreground and uniform priors for the signal parameters have been used. The authors have inferred tanh parameters from a fit to the reference ARES model. They have constructed synthetic datasets, by assuming an ideal instrument and observing four different sky regions. In later works, \citep{Mirocha_2015} have used broad, uninformative priors, but with simple foregrounds and ideal instruments. \citet{Bernardi_2016}, used uniform priors on all parameters. We have not made any such assumptions in our work, and have predicted the signal parameters with good accuracy. We do not require any specific model for fitting, neither do we require extensive prior information. So, we can consider this technique of signal extraction using ANN to be an alternative to the already existing techniques like MCMC explorations.\\
Our simple network has been able to detect the Global signal with an accuracy $\gtrsim 92 \%$, in cases of mock observations where the instrument has some residual time-varying gain across the spectrum. This is just a proof of concept paper, and we will deal with more complex scenarios in future work. The computations and results presented in this work has been performed on a personal computer (Macbook Air, 8GB RAM) only and does not require too much computational power. This shows how conveniently these algorithms can be used on any PC. However, in order to establish the robustness of the ANN framework for the Global 21cm signal extraction, we need to test this algorithm for more realistic instrument responses like that of EDGES, SARAS, etc. We also plan to address the issue of effect of chromatic primary beam and changes in the foreground model spectrum due to the same. In future, we would like to use ionospheric models to corrupt the datasets and extract 21cm signal in
presence of the atmosphere. Moreoever, the 21cm signal model is now dependent on `tanh' parametrization. But ANN can handle an non-parametrized 21cm signal. This will be used in future publications. At the end, we wish to apply the developed ANN framework on some real data for validation.

\newpage

\section{Acknowledgements}
The authors would like to thank the anonymous referee for insightful suggestions, which has considerably helped us to improve the manuscript. We thank Jordan Mirocha, Rohan Pattnaik, Avinash Deshpande and Suman Majumder for their help and fruitful discussions. We acknowledge the support of DST for providing the INSPIRE fellowship (IF160153). 




\bibliographystyle{mnras}
\bibliography{References} 








\bsp	
\label{lastpage}
\end{document}